%% file: StackedEDR.tex
\documentclass{aa}

\usepackage{graphicx}
\usepackage{txfonts}
\usepackage{hyperref}
\usepackage{amssymb}
\usepackage{amsmath}
\hypersetup{
    colorlinks=true,
    linkcolor=blue,
    filecolor=magenta,
    urlcolor=cyan,
    pdftitle={EDR VS sources},
    pdfpagemode=FullScreen,
    }

\input{Definitions.tex}
\def\myfig#1{Figures/#1}
\newcommand{\myP}{{\mathbb{P}}}
\newcommand{\NS}{\mathcal{N}}
\newcommand{\FS}{\mathcal{F}}
\newcommand{\Myc}{{\mathsf{c}}}

\defcitealias{reiss2018detection}{RK18}
\newcommand{\RK}{{\citetalias{reiss2018detection}}}
\defcitealias{HouEtAl23}{H23}
\newcommand{\HK}{{\citetalias{{HouEtAl23}}}}
\defcitealias{IlaniEtAl24}{I24}
\newcommand{\Ilani}{{\citetalias{{IlaniEtAl24}}}}

\begin{document}

   \title{Galaxy cluster virial-shock sources in eROSITA catalogs}
    %  \subtitle{Optional}

   \author{
   Gideon Ilani\thanks{Posthumously. Gideon Ilani was killed in action on December 10, 2023. This paper is based in part on his Ph.D. research.}
   \and
   Kuan-Chou Hou
   \and
   Gil Nadler
   \and
   Uri Keshet\thanks{\email{keshet.uri@gmail.com}}
   }

   \institute{Physics Department, Ben-Gurion University of the Negev, POB 653, Be'er-Sheva 84105, Israel}

   %\date{Received September xx, xxxx; accepted March xx, xxxx}
    \date{\today}

% \abstract{}{}{}{}{}
% 5 {} token are mandatory

  \abstract
  % Context: (optional, leave empty {} if necessary)
   {Virial shocks around galaxy clusters and groups are being mapped, tracing accretion onto large-scale structure.}
  % Aims:
   {Following the recent identification of discrete \emph{ROSAT} and radio sources associated with the virial shocks of MCXC clusters and groups, we examine if the early \emph{eROSITA-DE} data release (EDR) shows virial-shock X-ray sources within its $140$ deg$^2$ field.}
  % Methods:
   {EDR catalog sources are stacked and radially binned around EDR catalog clusters and groups.
   The properties of the excess virial-shock sources are inferred statistically by comparing the virial-shock region to the field.}
  % Results:
   {An excess of X-ray sources is found narrowly localized at the $2.0<r/R_{500}<2.25$ normalized radii, just inside the anticipated virial shocks, of the resolved 532 clusters, for samples of both extended ($3\sigma$ for 534 sources) or bright ($3.5\sigma$ for 5820 sources; $4\sigma$ excluding the low cluster-mass quartile) sources.
   The excess sources are on average extended ($\sim 100\kpc$), luminous ($L_X\simeq 10^{43\mbox{\scriptsize--}44}$ erg s$^{-1}$), and hot ($\sim$keV), consistent with infalling gaseous halos crossing the virial shock. The results agree with the stacked \emph{ROSAT}--MCXC signal, showing the higher $L_X$ anticipated at EDR redshifts and a possible dependence upon host mass.}
  % Conclusions: (optional, leave empty {} if necessary)
   {Localized virial-shock spikes in the distributions of discrete radio, X-ray, and probably also \gama-ray sources are new powerful probes of accretion from the cosmic web, with strong constraints anticipated with future all-sky catalogs such as by \emph{eROSITA}.}


   \maketitle
%
%-------------------------------------------------------------------

\section{Introduction}

In recent years, the long-awaited virial-shock (VS) signals around galaxy clusters and groups (for brevity, henceforth 'clusters') were finally detected in inverse-Compton \citep{KeshetEtAl17, ReissEtAl17, reiss2018detection, KeshetReiss18}, synchrotron \citep{KeshetEtAl17, HouEtAl23}, and thermal SZ \citep{KeshetEtAl17, HurierEtAl19, keshet20coincident, PrattEtAl21, Anbajagane2022} signatures, both in stacking analyses and in individual clusters.
The stacked leptonic signals indicate highly localized emission at normalized $2.2\lesssim\tau\equiv r/R_{500}\lesssim2.5$ radii, where subscript $500$ refers (henceforth) to the radius around a cluster enclosing $500$ times the critical mass density of the Universe.

An unexpected signal was reported recently \citep[][henceforth \Ilani]{IlaniEtAl24} in X-ray and radio catalog sources stacked around MCXC \citep{PiffarettiEtAl11} clusters, with a highly localized, $2.25<\tau<2.50$ excess precisely coincident with the previous VS leptonic signals.
These sources were found to be on average extended, $\sim$keV hot, magnetized, and radially polarized, and so were tentatively identified as the shocked halos of infalling galaxies or galaxy aggregates, possibly including aging relativistic particles from previous galactic outflows (\Ilani).
However, these stacking analyses of leptonic emission or discrete sources relied on the same low-redshift MCXC clusters and their tabulated, X-ray-based, characteristic $R_{500}$ values.

We examine if the early \emph{eROSITA-DE} data release (EDR) catalogs (described in  \S\ref{sec:Catalogs}) are sufficient to show an excess of VS X-ray sources within their $140$ deg$^2$ field, and, if so, to characterize the properties of these sources.
EDR catalog sources are thus stacked and radially binned around EDR catalog clusters (in  \S\ref{sec:Stacking}).
The properties of the excess sources are then inferred statistically (in \S\ref{sec:Properties}), by comparing VS-region sources to their field counterparts. Finally, the results are analyzed and discussed (in \S\ref{sec:Discussion}) in comparison to the \emph{ROSAT}--MCXC results.

We generally follow the {\Ilani} methods and notations.
A $\Lambda$CDM model is adopted with an $H_0=70\km\se^{-1}\Mpc^{-1}$ Hubble constant and an $\Omega_m=0.3$ mass fraction.

\section{Catalog samples}
\label{sec:Catalogs}

We combine the EDR\footnote{\href{https://erosita.mpe.mpg.de/edr/eROSITAObservations/Catalogues/}{\url{https://erosita.mpe.mpg.de/edr/eROSITAObservations/Catalogues/}}.} catalogs of X-ray sources
\citep{BrunnerEtAl22eRosita}, clusters \citep{LiuEtAl22eRosita}, and cluster X-ray properties \citep{BaharEtAl22eRosita}.
Better results are expected with the first \emph{eROSITA} allsky survey\footnote{\href{https://erosita.mpe.mpg.de/dr1/AllSkySurveyData_dr1/Catalogues_dr1/}{\url{https://erosita.mpe.mpg.de/dr1/AllSkySurveyData_dr1/Catalogues_dr1/}}.} (eRASS1) and future all-sky catalogs, as they become available\footnote{The cluster catalog for the first data release (DR1) of the SRG/eROSITA all-sky survey (eRASS1) has become available only after this study was concluded.}.

\begin{figure}[h]
    \centerline{\includegraphics[width=0.95\linewidth,trim={0cm 0cm 0cm 0cm},clip]{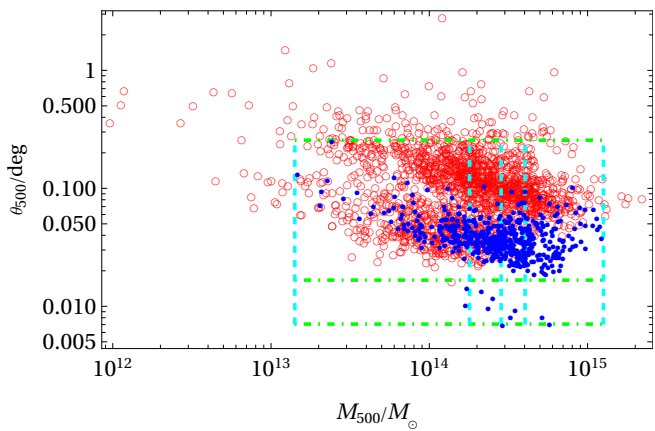}}
	\caption{\label{fig:Mvstheta}
      The $M_{500}$--$\theta_{500}$ phase space of \emph{eROSITA} EDR clusters (blue disks), shown in comparison to MCXC clusters (red circles), with boundaries demarking the four equal-sized mass bins (vertical dashed lines) and the ten excluded $\theta_{500}<1'$ clusters (horizontal dot-dashed).
    }
\end{figure}

Figure \ref{fig:Mvstheta} presents the EDR and MCXC cluster catalogs in the phase space of $M_{500}$ mass vs. projected $\theta_{500}=R_{500}/d_A$ angle, where $d_A(z)$ is the angular diameter distance at redshift $z$.
Thanks to the better resolution and sensitivity of \emph{eROSITA}, EDR clusters, including massive ones, are available at higher redshifts and thus smaller $\theta_{500}$ than in MCXC.
However, due to the smaller field of view, the EDR catalog lacks rare, highly extended or very massive clusters found in MCXC.
We divide the 542 EDR clusters into four mass bins, each with about the same, 135 or 136 number of clusters.
Due to the $\sim1'$ resolution of the cluster catalog \citep{LiuEtAl22eRosita}, we exclude the 10 clusters with $\theta_{500}<1'$.
The mass bins and $\theta_{500}$ cutoff are shown as lines in the figure.

\begin{figure}[h]
    \centerline{\includegraphics[width=0.95\linewidth,trim={0cm 0cm 0cm 0cm},clip]{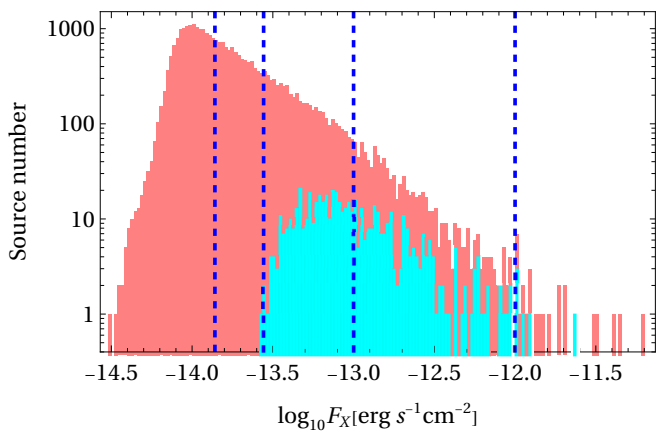}}
	\caption{\label{fig:Flux}
      Nominal ($[0.2,2.3]\keV$) flux distributions of extended (lower, cyan histogram) and all (higher, pink histogram) EDR catalog sources, with boundaries designating (vertical, left-to-right dashed lines) the full-catalog median, mean $F_{\rm mean}\simeq F_{\rm min}$, threshold $F_{\rm th}$, and upper cutoff $F_{\rm max}$ flux levels.
    }
\end{figure}

Figure \ref{fig:Flux} presents the X-ray $F_X$ flux histogram in the $[0.2,2.3]\keV$ single detection-band (henceforth, unless otherwise stated) among all 27910, likelihood $\mathcal{L}\geq6$ EDR catalog sources, and among the subset of 541 extended sources.
We focus on the regime between the minimal $F_{\rm min}\simeq 10^{-13.57}\erg\se^{-1}\cm^{-2}$ flux of the extended-source distribution, which approximately coincides with the mean flux $F_{\rm mean}$ of the full sample, and the upper cutoff $F_{\rm max}\simeq 10^{-12}\erg\se^{-1}\cm^{-2}$ imposed to avoid the apparent bright outliers, leaving 5820 sources, 534 of them extended.
We further separate this range into faint vs. bright sources at a threshold $F_{\rm th}\simeq 10^{-13}\erg\se^{-1}\cm^{-2}$, approximately coinciding with the extended-source median. These flux levels, as well as the median of the full sample, are presented as vertical dashed lines in the figure.
The $\sim10''$ resolution of the source catalog \citep[adopting the camera pixel size;][]{BrunnerEtAl22eRosita} is sufficient for the analysis of $\theta_{500}\gtrsim1'$ clusters.

\section{Binning and Stacking}
\label{sec:Stacking}

Denoting $\NS(\tau,\Myc)$ as the number of sources found in a radial ring of radius $\tau$ and width $\Delta\tau$ around cluster $1\leq\Myc\leq N_c$, and $\FS(\tau,\Myc)$ as the number of combined foreground and background (henceforth referred to as field) sources anticipated in this ring, we adopt the field-only, $\NS=\FS$ null hypothesis.
A positive $\NS>\FS$ excess is assigned a source-weighted (SW)
\begin{equation}\label{eq:SigSW}
 S_{\rm SW}(\tau) = \frac{\NS(\tau)-\FS(\tau)}{\sqrt{\FS(\tau)}}
 \,\,\, \biggm\lvert \,\,\, \{\NS,\FS\}(\tau)\equiv\sum\limits_{\Myc=1}^{N_c}\{\NS,\FS\}(\tau,\Myc)\, ,
\end{equation}
or a cluster-weighted (CW)
\begin{equation}\label{eq:SigSW}
 S_{\rm CW}(\tau) = \frac{1}{\sqrt{N_c}}\sum\limits_{\Myc=1}^{N_c} \frac{\NS(\tau,\Myc)-\FS(\tau,\Myc)}{\sqrt{\FS(\tau,\Myc)}}\fin
\end{equation}
significance in the $\FS\gg1$, normal-distribution limit.
Modification for a Poisson distribution, necessary for finite and especially small $\FS$, are provided in \S\ref{app:equations} and incorporated henceforth.

For simplicity, given the EDR limited field of view, intermediate, $20\dgr\lesssim b\lesssim 40\dgr$ Galactic latitudes, and large exposure variations especially in the field periphery \citep{BrunnerEtAl22eRosita}, we approximate $\FS$ as a constant, measured in the $\{130\dgr<\mbox{RA}<142\dgr$, $0\dgr<\mbox{Dec}<4\dgr\}$ rectangle of fairly uniform exposure. The results are not sensitive to reasonable changes in the determination of $\FS(\tau)$, including field estimates in the vicinity of each cluster and polynomial sky fits (see \Ilani).
We use the same nominal $\Delta\tau=0.25$ resolution used in previous stacking analyses.

\begin{figure*}[h]
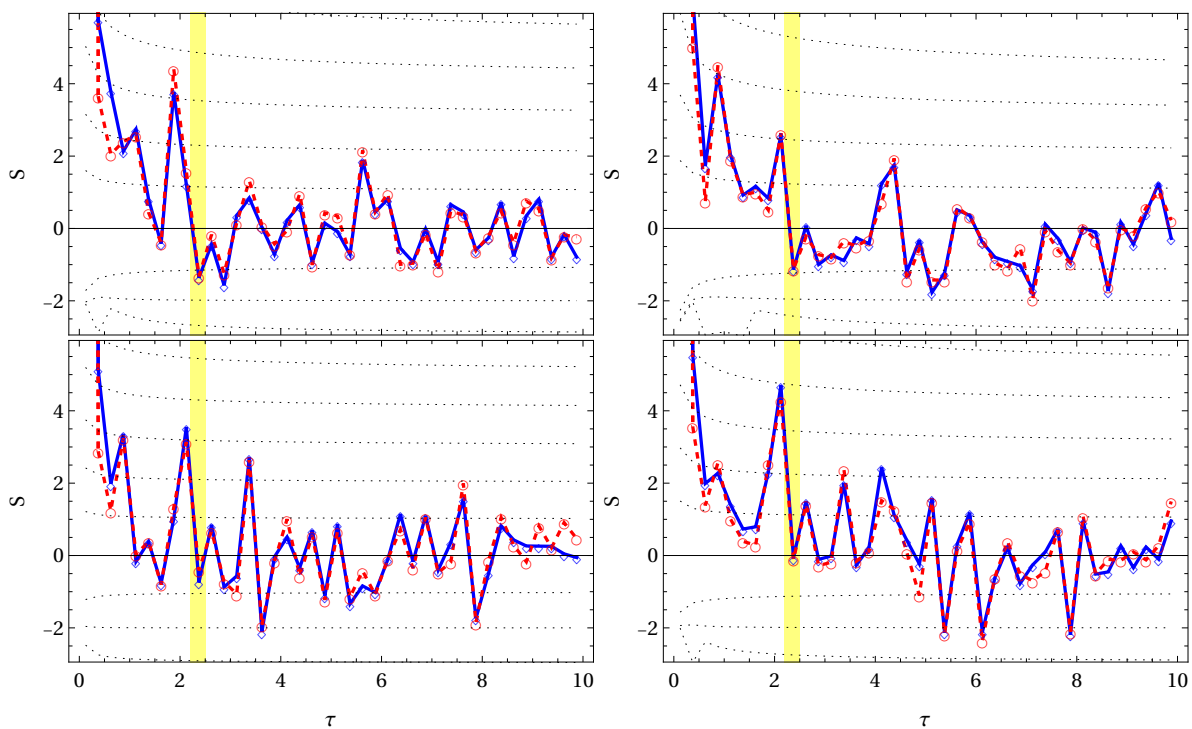

    \centerline{
        \includegraphics[width=0.42\linewidth,trim={0cm 1cm 0cm 0cm},clip]{\myfig{ExtM234Fmean.eps}}
        \includegraphics[width=0.42\linewidth,trim={0cm 1cm 0cm 0cm},clip]{\myfig{ExtM234Fhigh.eps}}
    }
    \centerline{
        \includegraphics[width=0.42\linewidth,trim={0cm 0cm 0cm 0cm},clip]{\myfig{StackM234Fmean.eps}}
        \includegraphics[width=0.42\linewidth,trim={0cm 0cm 0cm 0cm},clip]{\myfig{StackM234Fhigh.eps}}
    }
	\caption{\label{fig:StackDemo}
      Top row: significance $S(\tau)$ radial profiles of all (left panel) or only bright $F>F_{\rm th}$ (right) extended \emph{eROSITA} sources, SW (blue diamonds with solid lines to guide the eye) or CW (red circles with dashed guide) stacked around massive (bins 2--4) clusters. Also shown are Poisson-statistics confidence levels ($\pm\{1\sigma,2\sigma,3\sigma,\ldots\}$; dotted curves) and the anticipated $2.2<\tau<2.5$ VS region (vertical yellow shading) based on previous stacked \gama-ray (\RK) and radio (\HK) continuum, and discrete X-ray and radio source (\Ilani), detections.
      Bottom row: same, but also including non-extended sources in the same $F_{\rm min}<F<F_{\rm max}$ range.
    }
\end{figure*}

Away from the central, $\tau\lesssim1$ excess of sources associated with the intracluster medium (ICM), the extended (extended and bright) sources present a $>3\sigma$ ($>2\sigma$) excess inside the anticipated VS radius, peaked at $1.75<\tau<2.0$ ($2.0<\tau<2.25$), as shown in Fig.~\ref{fig:StackDemo} (top row).
Incorporating also non-extended EDR sources yields a local $2.0<\tau<2.25$ excess in each mass bin 1--4, although the low mass bin 1 shows only a $\sim0.5\sigma$ excess, as compared to $S>2$ in each of the more massive bins 2--4; the co-added excess over these three bins, shown in Fig.~\ref{fig:StackDemo} (bottom row), presents a $\sim3\sigma$ ($\sim4\sigma$) excess for all (for bright) sources.
Results for each mass bin and for all bins combined are shown in \S\ref{app:mass}.
Note that while the VS signal emerges in extended or bright sources without any cuts, it vanishes if $F_{min}$ is lowered \eg to the catalog median, which is sensitive to the $\mathcal{L}$ cut.

The peripheral location of the excess and its proximity to the previously stacked signals tie it to the VS region.
The narrowness of the signal indicates that the excess sources are directly associated with the shock, rather than being driven by the gradual changes in environment as infalling objects approach the cluster.
The same findings in the \emph{ROSAT}--MCXC analysis, along with the properties of the excess X-ray and radio sources, led to the conclusion that these objects are likely shocked infalling gaseous halos, probably around galaxies or galaxy aggregates (\Ilani).
The VS excess typically peaks in the $2.0<\tau<2.25$ bin for EDR stacking, just inside the $2.25<\tau<2.5$ bin of maximal MCXC-stacking excess.
This small offset might arise from differences in the catalog prescriptions for $R_{500}$ determination, although some redshift dependence cannot be ruled out.

\section{Excess Source Properties}
\label{sec:Properties}

Although the localized $2.0<\tau<2.25$ (henceforth: the VS bin) excess is significant, the number of excess sources is small compared to the coincident field sources, so we are unable to tie individual sources to the VS. We can, however, characterize the excess sources statistically, by comparing catalog properties in the VS bin and in the field, although the sample is small.

Consider some source property $\myP$ of interest, such as source extent $\mbox{SE}$ or luminosity $L$. In the absence of source redshifts, we estimate $\myP$ by assigning sources with the redshifts of their tentative host clusters.
We then compare the differential $\mathbb{N}(\myP)\equiv dN/d\myP$ and cumulative $N(<\myP)$ distributions among the $\NS(2.0<\tau<2.25,\Myc)$ sources in the VS bin (subscript $V$) to the distributions of many random samples, each of $\FS(2.0<\tau<2.25,\Myc)$ sources, drawn from the field region (subscript $F$).
The cumulative VS excess $N_{\mbox{\tiny V}}(<\myP)-N_{\mbox{\tiny F}}(<\myP)$ and its local $S(\myP)=[\mathbb{N}_{\mbox{\tiny V}}-\mu(\mathbb{N}_{\mbox{\tiny F}})]/\sigma(\mathbb{N}_{\mbox{\tiny F}})$ significance are then estimated.
We also examine a reference ICM region, $1.0<\tau<1.5$, rescaling its $\NS$ to the solid angle of the VS bin.

\begin{figure}[b!]
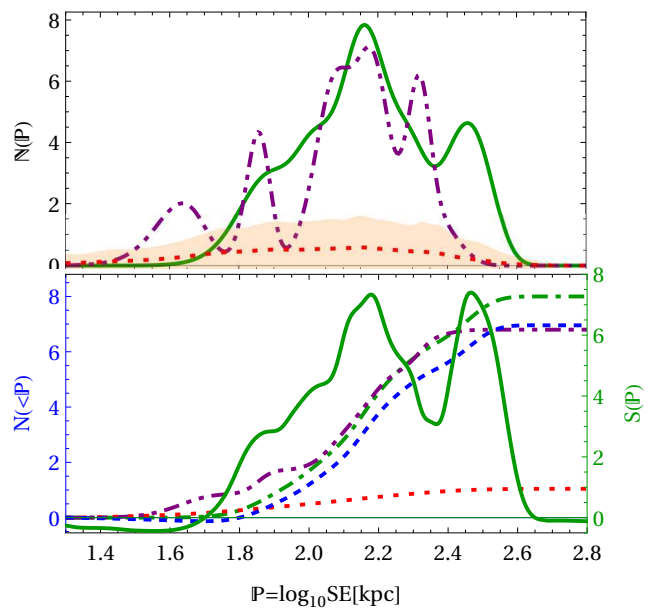

    \vspace{-0.07cm}
    \centerline{\includegraphics[width=0.922\linewidth,trim={0cm 1.02cm 0cm 0cm},clip]{\myfig{PropMFExt1.eps}}}
    \vspace{-0.07cm}
    \centerline{\hspace{-0.02cm}\includegraphics[width=0.92\linewidth,trim={0cm 0cm 0cm 0cm},clip]{\myfig{PropMFExt2.eps}}}
	\caption{\label{fig:MFSourceR}
      Source extent distribution for nominal ($F>F_{\rm th}$ and mass bins 2--4) cuts.
      Differential $\mathbb{N}(\myP)$ (top panel) and cumulative $N(<\myP)$ (bottom panel with left axis) distributions shown for $\myP=\log_{10}\mbox{SE}$ in the VS bin (solid and dot-dashed green), ICM region (double dot-dashed purple), field sampling (dotted red; pink-shaded $1\sigma$ dispersion), and VS excess (dashed blue). The bottom panel also shows the local significance $S(\myP)$ of the VS excess (solid green curve with right axis). See text.
    }
\end{figure}

\begin{figure}[h!]
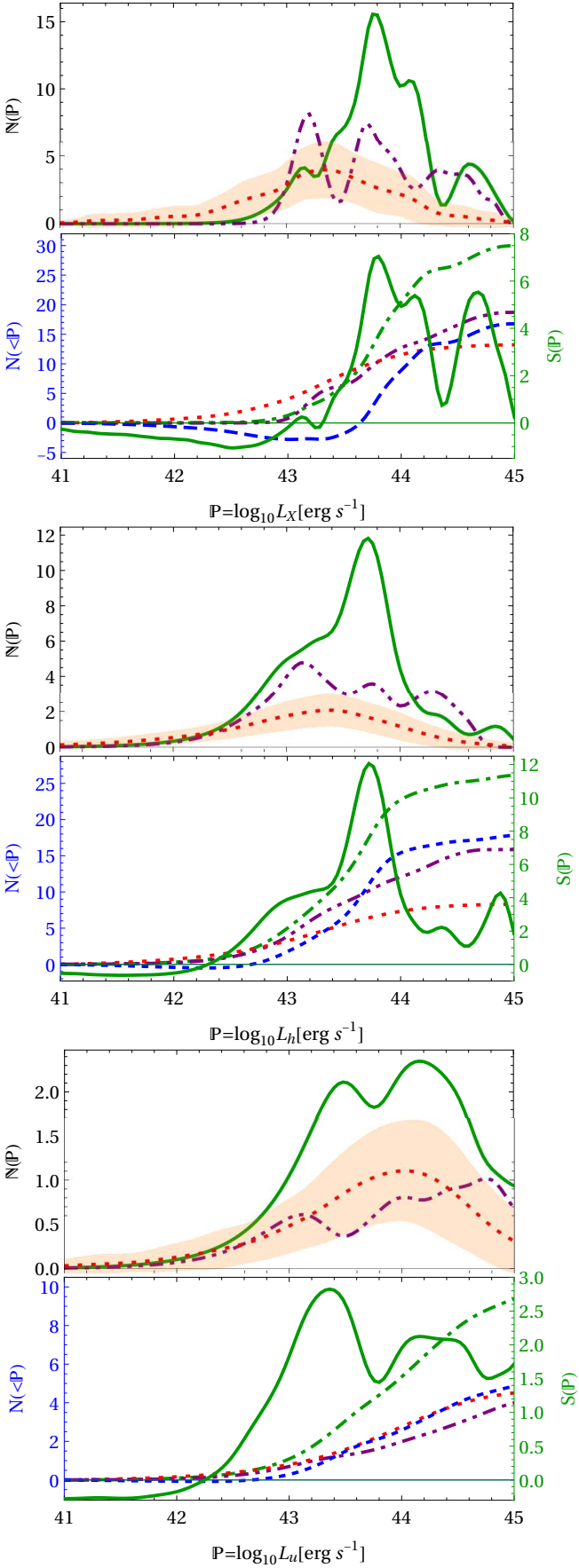

    \vspace{-0.07cm}
    \centerline{\includegraphics[width=0.923\linewidth,trim={0cm 0.9cm 0cm 0cm},clip]{\myfig{PropMFLum1.eps}}}
    \vspace{-0.15cm}
    \centerline{\includegraphics[width=0.92\linewidth,trim={0cm 0cm 0cm 0cm},clip]{\myfig{PropMFLum2.eps}}}
    \centerline{\includegraphics[width=0.923\linewidth,trim={0cm 0.9cm 0cm 0cm},clip]{\myfig{PropMFLumh1.eps}}}
    \vspace{-0.07cm}
    \centerline{\hspace{0.15cm}\includegraphics[width=0.94\linewidth,trim={0cm 0cm 0cm 0cm},clip]{\myfig{PropMFLumh2.eps}}}
    \centerline{\includegraphics[width=0.923\linewidth,trim={0cm 0.9cm 0cm 0cm},clip]{\myfig{PropMFLumu1.eps}}}
    \vspace{-0.17cm}
    \centerline{\hspace{0.28cm}\includegraphics[width=0.94\linewidth,trim={0cm 0cm 0cm 0cm},clip]{\myfig{PropMFLumu2.eps}}}
	\caption{\label{fig:MFSourceL}
      Luminosity $\myP=\log_{10}L$ distributions in the $[0.2,2.3]$ (top two panels), $[2.3,5]$ (middle panels), and $[5,8]$ keV (bottom panels) bands, using the same sample and notations of Fig.~\ref{fig:MFSourceR}.
      We add a $\sigma_{\rm smooth}^2 \equiv (0.2\,L)^2$ smoothing variance to $L$ uncertainties for visibility.
    }
\vspace{-1.cm}
\end{figure}

Figure \ref{fig:MFSourceR} presents the $\mbox{SE}$ distribution in the VS bin (green, solid and dot-dashed curves), as compared to the field (dotted red) and ICM (double-dot dashed purple) regions.
As the figure shows, there is a significant, $>3\sigma$ local excess of $\sim7$ sources with $60\kpc\lesssim\mbox{SE}\lesssim300\kpc$, broadly consistent\footnote{The $\mbox{SE}$, defined as the core radius in a $\beta=2/3$ model \citep{BrunnerEtAl22eRosita}, cannot be directly compared to the \emph{ROSAT} source radii (\Ilani).} with {\Ilani}.
The ICM region shows comparable or slightly smaller SE, similarly based on only a few sources.
The results in this section are based on bright, $F>F_{\rm th}$ sources around clusters in mass bins 2--4 (showing the most significant Fig.~\ref{fig:StackDemo} excess), but similar results are obtained for the entire sample with all mass bins; see \S\ref{app:MoreProps}.

Low-redshift, $L_X\sim 10^{42\mbox{--}43} \erg \se^{-1}$ VS-sources were identified by {\Ilani} as thermal emission from shocked infalling halos, so one may naively expect $L_X\propto (1+z)^6$ luminosities $\sim10$ times higher at typical EDR redshifts, in both nominal and higher energy bands.
Figure \ref{fig:MFSourceL} presents the logarithmic distributions of $L_X$, $L_h$, and $L_u$ luminosities, respectively, in the observed $[0.2,2.3]$, $[2.3,5]$, and $[5,8]$ keV bands.
A significant VS excess of $\sim15$ sources in the $10^{43\mbox{--}44}\erg\se^{-1}$ range is indeed found in $L_X$ and in $L_h$, whereas no significant excess can be identified in $L_u$.
These sources are thus consistent with thermal emission from gaseous objects compressed and heated to $\sim$keV temperatures by the VS; see also \S\ref{app:MoreProps}.

\section{Summary and Discussion}
\label{sec:Discussion}

Stacking EDR sources (Fig.~\ref{fig:Flux}) around 532 EDR clusters (Fig.~\ref{fig:Mvstheta}) shows a significant excess of extended ($3\sigma$) or bright ($3.5\sigma$) sources, narrowly localized at normalized $2.0<\tau<2.25$ radii (Figs.~\ref{fig:StackDemo}, \ref{fig:StackMbins}), like the \emph{ROSAT}--MCXC signal ({\Ilani}), thus directly linked to the VS shock of the host cluster.
The sources are on average extended (Figs.~\ref{fig:MFSourceR}, \ref{fig:SourceR}), with $L_X\simeq L_h\simeq 10^{43\mbox{--}44}\erg\se^{-1}$ about 10 times higher than in {\Ilani} sources, and no $L_u$ detection (Figs.~\ref{fig:MFSourceL}, \ref{fig:SourceL}), as expected for $\sim$keV sources at EDR redshifts.
The signal strengthens ($4\sigma$) without the low, $M_{500}<10^{14.3}$ mass quartile, suggesting a host-mass dependence not seen by {\Ilani}.

The stacked signals emerge thanks to the omission of faint, $F<F_{\rm min}\simeq F_{\rm mean}$ background sources (detected at lower likelihood levels), and the approximate similarity of clusters when scaled by $R_{500}$.
The small offset of the $2.0<\tau<2.25$ EDR signal from previous $2.2<\tau<2.5$ MCXC-based VS signals may hint at some dependence on redshift or $F$ cut, but may also arise from differences in the catalog prescriptions for $R_{500}$ determination, based on X-rays in MCXC vs. weak lensing in EDR.
Larger future studies, in particular using \emph{eROSITA} all-sky catalogs, could better resolve the offset and identify its origin, constrain $\mbox{SE}$, $L$, and additional source properties, test the tentative $M_{500}$ dependence, and examine deviations from spherical symmetry, which were suggested by {\Ilani} and earlier MCXC-based studies but are too weak for the EDR.

Our results are consistent with the tentative identification of the X-ray sources as thermal emission from gaseous objects, likely galactic halos, compressed and heated to $\sim$keV temperatures by the VS, with a strong $L_X(z)$ dependence.
Their non-thermal counterpart, already seen as excess synchrotron radio sources, should also have nonthermal hard X-ray and probably also \gama-ray counterparts, so combining or cross-correlating broadband catalogs should uncover more VS-related phenomena (\Ilani).
We conclude that localized VS spikes in the distributions of discrete sources are a powerful probe of accretion from the cosmic web and of the underlying physical processes, ranging from the evolution of large-scale structure, to magnetization by the VS, to its dark-matter splashback counterpart.

\begin{acknowledgements}
This research received funding from ISF grant No. 2126/22. \\
May Gideon Ilani's memory be a blessing.
\end{acknowledgements}

\bibliographystyle{aa}
\bibliography{Virial}

\appendix

\section{Stacking equations}
\label{app:equations}

For a Poisson distribution of mean $\lambda$, where measurement $k$ has probability $p(k)= e^{-\lambda} \lambda^k/k!$, we may analytically sum the probabilities of equal or larger (smaller) $k$ values to quantify the significance of a positive excess (negative deficit) in units of Gaussian standard errors,
\begin{equation}
\label{eq:PoissonCL_corrected0}
S_p(k;\lambda) =
\begin{cases}
\vspace{0.1cm}
\sqrt{2}\,\mathrm{erfc}^{-1}\left[\frac{\Gamma(k)-\Gamma(k,\lambda)}{\Gamma(k)/2}\right]>0 & \mbox{if } k>\lambda\,; \\
\vspace{0.1cm}
-\sqrt{2}\,\mathrm{erfc}^{-1}\left[\frac{\Gamma(1+k,\lambda)}{\Gamma(1+k)/2}\right]<0 & \mbox{if } k<\lambda\,; \\
0 & \mbox{if } k=\lambda \,,
\end{cases}
\end{equation}
where $\mathrm{erfc}(x)$ and $\Gamma(x)$ are respectively the complementary error and Euler Gamma functions.
One can find $S_{\rm SW}$ for given $S_p$ and $\FS$ by solving (\emph{cf.} Eq.~\ref{eq:SigSW})
\begin{equation}\label{eq:PoissonCL_corrected1}
  S_p = S_p(\NS=\FS+S_{\rm SW}\sqrt{\FS};\FS) \coma
\end{equation}
resulting in the $S_{\rm SW}(\tau)$ profiles shown (dotted curves) for integer $S_p$ values in the $S(\tau)$ figures \ref{fig:StackDemo} and \ref{fig:StackMbins}.

\section{Source properties in the full sample}
\label{app:MoreProps}

\begin{figure}[h!]
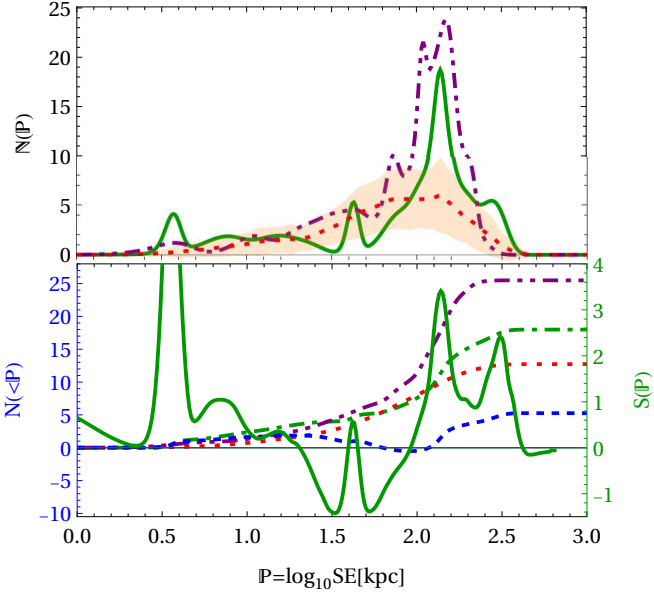

    \vspace{-0.07cm}
    \centerline{\hspace{0.02cm}\includegraphics[width=0.922\linewidth,trim={0cm 0.9cm 0cm 0cm},clip]{\myfig{PropExt1.eps}}}
    \vspace{-0.15cm}
    \centerline{\includegraphics[width=0.95\linewidth,trim={0cm 0cm 0cm 0cm},clip]{\myfig{PropExt2.eps}}}
	\caption{\label{fig:SourceR}
      Source extent distribution among all sample sources around all clusters (same notations as in Fig.~\ref{fig:MFSourceR}).
    }
\end{figure}

\begin{figure}[h!]
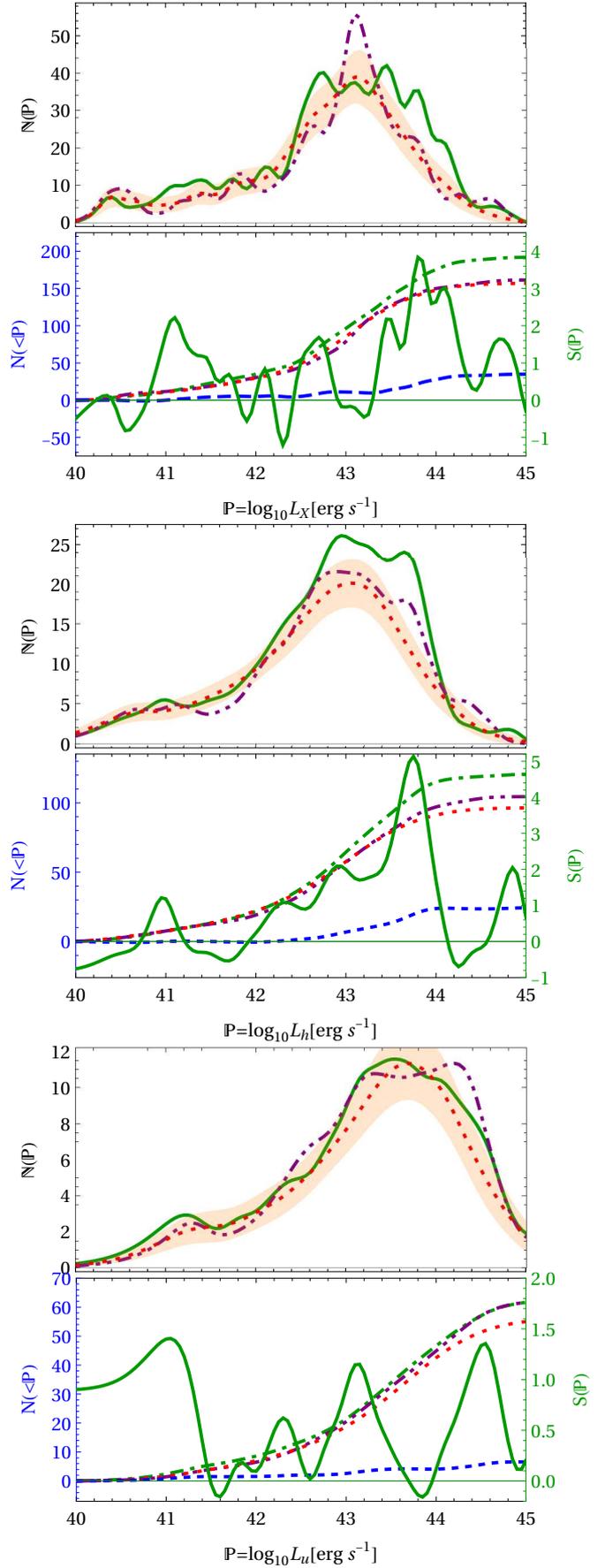

    \vspace{-0.07cm}
    \centerline{\includegraphics[width=0.923\linewidth,trim={0cm 0.9cm 0cm 0cm},clip]{\myfig{PropLum1.eps}}}
    \vspace{-0.05cm}
    \centerline{\hspace{-0.03cm}\includegraphics[width=0.95\linewidth,trim={0cm 0cm 0cm 0cm},clip]{\myfig{PropLum2.eps}}}
    \centerline{\includegraphics[width=0.923\linewidth,trim={0cm 0.9cm 0cm 0cm},clip]{\myfig{PropLumh1.eps}}}
    \vspace{-0.05cm}
    \centerline{\hspace{-0.03cm}\includegraphics[width=0.95\linewidth,trim={0cm 0cm 0cm 0cm},clip]{\myfig{PropLumh2.eps}}}
    \centerline{\includegraphics[width=0.923\linewidth,trim={0cm 0.9cm 0cm 0cm},clip]{\myfig{PropLumu1.eps}}}
    \vspace{-0.15cm}
    \centerline{\hspace{0.22cm}\includegraphics[width=0.945\linewidth,trim={0cm 0cm 0cm 0cm},clip]{\myfig{PropLumu2.eps}}}
	\caption{\label{fig:SourceL}
    Luminosity distributions among all sample sources around all clusters (same notations as in Fig.~\ref{fig:MFSourceL}).
    }
\vspace{-0.8cm}
\end{figure}

Figures \ref{fig:SourceR} and \ref{fig:SourceL} complement Figs. \ref{fig:MFSourceR} and \ref{fig:MFSourceL}, respectively, showing the properties of all $F_{\rm min}<F<F_{\rm max}$ sources around clusters in all mass bins.
The signals are stronger here, \ie they involve more excess sources than in Figs. \ref{fig:MFSourceR} and \ref{fig:MFSourceL}, but are more noisy; the conclusions are qualitatively unchanged.

\section{Mass dependence}
\label{app:mass}

Figure \ref{fig:StackMbins} presents the stacked $S(\tau)$ profile in each of the four mass bins shown in Fig.~\ref{fig:Mvstheta}, as well as the signal co-added over all four mass bins.
While the higher mass bins 2--4 each presents a $\sim2\sigma$ excess in the VS bin, the $M\lesssim 10^{14.3}$ clusters in mass bin 1 show a negligible excess.
This result differs from the comparable signals found by {\Ilani} in all their mass bins; however, the poor statistics of the present sample cannot substantiate a meaningful difference between EDR and \emph{ROSAT}--MCXC results.

\def\FigSize{0.37}
\begin{figure*}[h]
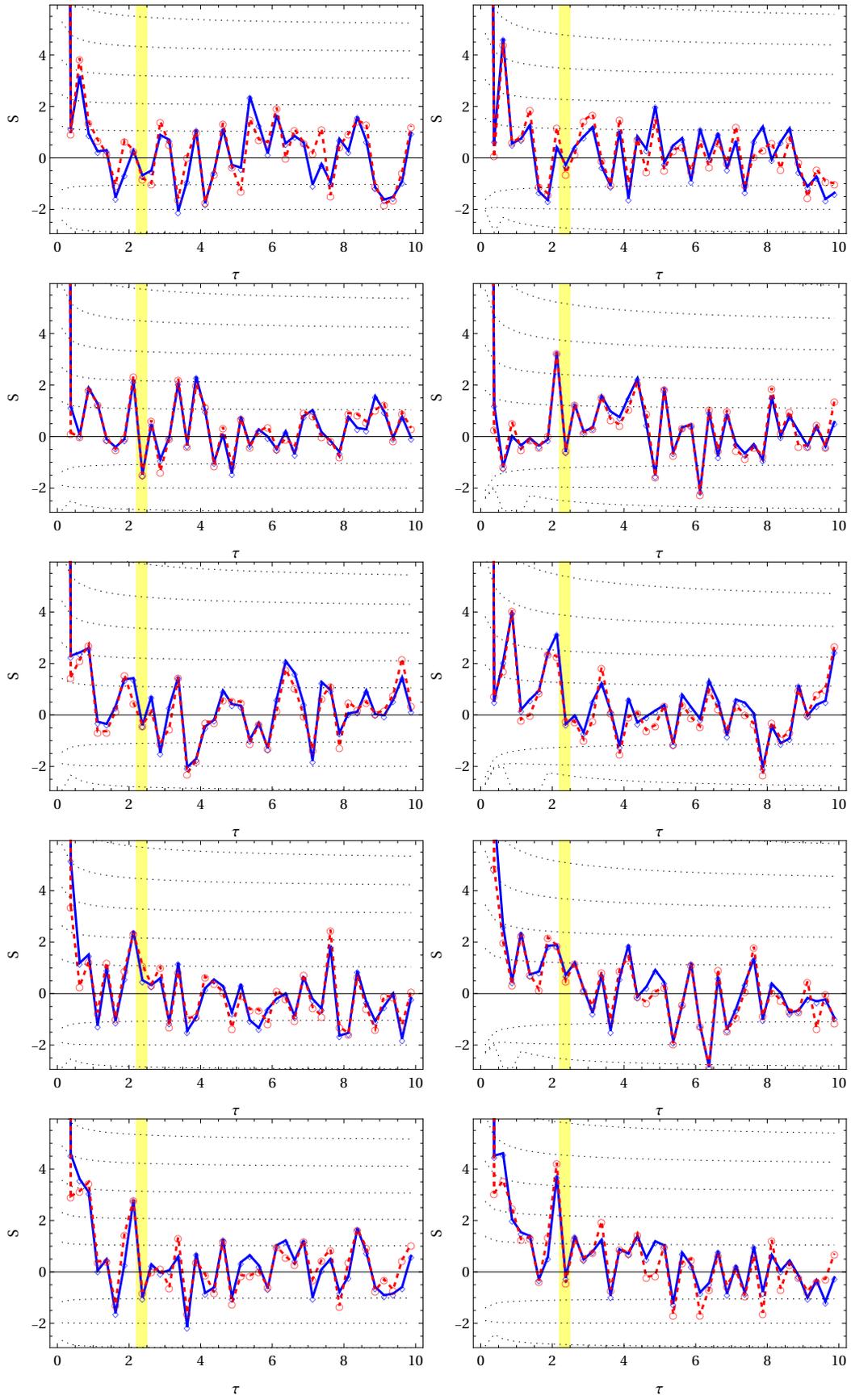

    \centerline{
        \includegraphics[width=\FigSize\linewidth,trim={0cm 0cm 0cm 0cm},clip]{\myfig{StackM1Fmean.eps}}
        \includegraphics[width=\FigSize\linewidth,trim={0cm 0cm 0cm 0cm},clip]{\myfig{StackM1Fhigh.eps}}
    }
    \centerline{
        \includegraphics[width=\FigSize\linewidth,trim={0cm 0cm 0cm 0cm},clip]{\myfig{StackM2Fmean.eps}}
        \includegraphics[width=\FigSize\linewidth,trim={0cm 0cm 0cm 0cm},clip]{\myfig{StackM2Fhigh.eps}}
    }
    \centerline{
        \includegraphics[width=\FigSize\linewidth,trim={0cm 0cm 0cm 0cm},clip]{\myfig{StackM3Fmean.eps}}
        \includegraphics[width=\FigSize\linewidth,trim={0cm 0cm 0cm 0cm},clip]{\myfig{StackM3Fhigh.eps}}
    }
    \centerline{
        \includegraphics[width=\FigSize\linewidth,trim={0cm 0cm 0cm 0cm},clip]{\myfig{StackM4Fmean.eps}}
        \includegraphics[width=\FigSize\linewidth,trim={0cm 0cm 0cm 0cm},clip]{\myfig{StackM4Fhigh.eps}}
    }
    \centerline{
        \includegraphics[width=\FigSize\linewidth,trim={0cm 0cm 0cm 0cm},clip]{\myfig{StackMAllFmean.eps}}
        \includegraphics[width=\FigSize\linewidth,trim={0cm 0cm 0cm 0cm},clip]{\myfig{StackMAllFhigh.eps}}
    }
	\caption{\label{fig:StackMbins}
        Same as Fig.~\ref{fig:StackDemo}, but showing mass bins 1--4 separately (first four panels, top down) and combined (bottom panel).
    }
\end{figure*}

\end{document}

%% file: Definitions.tex
\newcommand{\ie}{\emph{i.e.} }
\newcommand{\eg}{\emph{e.g.,} }

\newcommand{\be}{\begin{equation}}
\newcommand{\ee}{\end{equation}}
\newcommand{\bea}{\begin{equation*}}
\newcommand{\eea}{\end{equation*}}
\newcommand{\beq}{\begin{equation} }
\newcommand{\eeq}{\end{equation}}
\newcommand{\beqr}{\begin{eqnarray} \nonumber}
\newcommand{\eeqr}{\end{eqnarray}}
\newcommand{\beqrb}{\begin{eqnarray}}
\newcommand{\eeqrb}{\nonumber \end{eqnarray}}
\newcommand{\fin}{\mbox{ .}}
\newcommand{\coma}{\mbox{ ,}}

\newcommand{\cm}{\mbox{ cm}}

\newcommand{\se}{\mbox{ s}}

\newcommand{\erg}{\mbox{ erg}}

\newcommand{\km}{\mbox{ km}}

\newcommand{\kpc}{\mbox{ kpc}}
\newcommand{\Mpc}{\mbox{ Mpc}}

\newcommand{\keV}{\mbox{ keV}}

\newcommand{\gama}{$\gamma$}

\newcommand{\dgr}{^{\circ}}